\documentclass{revtex4-2}
\usepackage[utf8]{inputenc}
\usepackage{amsbsy}
\usepackage{amsopn}
\usepackage{amstext}
\usepackage{amsmath,amsthm,amsfonts,amssymb}
\usepackage[mathcal]{eucal}
\usepackage{mathrsfs}
\usepackage{braket}
\usepackage[english]{babel}
\usepackage{color}
\usepackage{esint}
\usepackage{graphicx}
\usepackage{float}
\usepackage{units}
\usepackage{blindtext}
\usepackage{hyperref}
\usepackage{textcomp}
\usepackage{orcidlink}

\DeclareGraphicsExtensions{.png,PNG,.pdf,.PDF}
\usepackage{hyperref}
\usepackage{slashed}
\newcommand{\ie}{\begin{equation}}
\newcommand{\fe}{\end{equation}}
\newcommand{\se}{\begin{eqnarray}}
\newcommand{\ff}{\end{eqnarray}}
\begin{document}

\title{Comment on ``Unruh effect for neutrinos interacting with accelerated matter''}
\author{R. R. S. Oliveira\,\orcidlink{0000-0002-6346-0720}}
\email{rubensrso@fisica.ufc.br}
\affiliation{Departamento de F\'isica, Universidade Federal do Cear\'a (UFC), Campus do Pici, C.P. 6030, Fortaleza, CE, 60455-760, Brazil}


\date{\today}

\begin{abstract}

In the present comment, we show that the fundamental equation worked by Dvornikov in his paper, which is the Dirac equation for a massive neutrino interacting with linearly accelerated matter, is incorrect. In particular, Dvornikov incorrectly wrote/defined the effective external current in a curved space-time. In other words, Dvornikov wrote/defined such an effective current in a flat space-time, which is a mistake. Consequently, the second-order differential equation (generated through the quadratic Dirac equation) in your paper is incorrect, where such an equation is given by the Whittaker equation. So, since the solutions of such a differential equation (whose solutions are the Whittaker functions) are the basis for its results, it implies that such results are also incorrect. In this way, starting from the true/correct Dirac equation with an effective external current in a curved space-time, we obtain in detail the second-order differential equation (also a Whittaker equation) and its solutions for a neutrino interacting with linearly accelerated matter.

\end{abstract}

\maketitle

\section{Introduction}

In a paper published in the Journal of High Energy Physics (JHEP), entitled ``Unruh effect for neutrinos interacting with accelerated matter'', Dvornikov \cite{Dvornikov} studied the evolution of neutrinos in a background matter moving with a linear acceleration (i.e., Rindler space-time). To do such a study, Dvornikov \cite{Dvornikov} worked with the Dirac equation for a massive neutrino electroweakly interacting with background fermions. In other words, Dvornikov \cite{Dvornikov} worked with the Dirac equation in a curved space-time in three-dimensional Cartesian coordinates and subject to an effective external current of background fermions. In fact, this was done because the curved Dirac equation (written in terms of the vierbein vectors) can be used perfectly to introduce the non-inertial effects of a linear acceleration (or uniform rotation) into the system (i.e., this is a consequence of Einstein's equivalence principle of general relativity, which states that the effects of gravity are (locally) indistinguishable from the effects of uniform acceleration). So, once this Dirac equation (actually a second-order differential equation) was solved for ultrarelativistic neutrinos ($m\to 0$), Dvornikov \cite{Dvornikov} obtained the neutrino quantum states. Next, Dvornikov \cite{Dvornikov} demonstrated that the neutrino electroweak interaction with accelerated matter leads to the vacuum instability, which results in the neutrino-antineutrino pairs creation. Also, Dvornikov \cite{Dvornikov} rederived the temperature of the Unruh radiation and found the correction to the Unruh effect due to the neutrino interaction with background fermions. In particular, this paper is well-written and covers a very interesting topic about Dirac neutrinos interacting with matter in an accelerated frame. As for the formalism used by Dvornikov \cite{Dvornikov} (vierbein vectors/spin connection formalism of general relativity), it is also very important in the literature when working with the Dirac equation in curved space-times or with non-inertial effects \cite{Carvalho,Cuzinatto,Bragança,Chen,Guvendi,Vitoria,Bakke,Cunha,O1,O2,O3,O4,O5,O6,JHEP,JCAP,Bandyopadhyay,JHEP2,JCAP2}.

However, according to two recently published papers on neutrinos in curved space-times and with rotation effects (one published in Physical Review D (PRD) \cite{Bandyopadhyay} and the other in the Journal of Cosmology and Astroparticle Physics (JCAP) \cite{JCAP}), we note that the Dirac equation worked by Dvornikov \cite{Dvornikov} in his paper is incorrect. That is, the Dirac equation for a neutrino interacting with accelerated matter was written/defined incorrectly in a curved space-time. In particular, Dvornikov \cite{Dvornikov} incorrectly wrote/defined the effective external current of background fermions in a curved space-time (in other words, he wrote/defined such a current in a flat space-time, which is a mistake). Consequently, the second-order differential equation (Whittaker equation) in your paper is incorrect. So, since the solutions of such a differential equation (whose solutions are the Whittaker functions) are the basis for its results, it implies that such results are also incorrect. In addition to this error, another error (actually a contradiction) committed by Dvornikov \cite{Dvornikov} (also found in \cite{JCAP2}) was to consider the neutrino mass in the results ($m\neq 0$), even though he claimed to have worked with ultrarelativistic neutrinos whose mass is negligible ($m\to 0$). That is, Dvornikov \cite{Dvornikov} stated that his differential equation could only be solved for $m\to 0$; however, he still considered $m\neq 0$. Therefore, based on Refs. \cite{Bandyopadhyay,JCAP} (and mainly \cite{JCAP}, as well as \cite{Bragança}), the present comment has as its goal to obtain in detail the true/correct second-order differential equation (another Whittaker equation) generated from the curved Dirac equation for a neutrino interacting with linearly accelerated matter (with $m\neq 0$). Besides, we will also obtain the true/correct Whittaker functions.


\section{Quick review of the main steps that Dvornikov took to obtain the second-order differential equation from the curved Dirac equation for a neutrino interacting with linearly accelerated matter}

According to Dvornikov \cite{Dvornikov}, the Dirac equation for a neutrino interacting with background matter in a curved space-time is written in the form
\begin{equation}\label{1}
[i\gamma^\mu (x)\nabla_\mu (x)-m]\psi=\frac{1}{2}J^\mu\gamma_\mu (x)[1-\gamma^5 (x)]\psi, \ \ (\mu=t,x,y,z),
\end{equation}
where $\psi$ is the neutrino bispinor, $m$ is the neutrino mass, $\gamma^\mu (x)=e^{\ \mu}_a (x)\gamma^a$ and $\gamma_\mu (x)=e^a_{\ \mu}(x)\gamma_a$ ($a=0,1,2,3$) are the curved gamma matrices (or coordinate dependent Dirac matrices), $\gamma^a$ and $\gamma_a$ are the flat gamma matrices or standard/usual Dirac matrices (with $\gamma^0=\gamma_0$ and $\gamma^i=-\gamma_i$; $i=1,2,3$), $e^{\ \mu}_a(x)$ are the vierbein vectors and $e^a_{\ \mu}(x)$ are their inverse (and satisfy the orthogonality condition given by $e^a_{\ \mu}(x) e^{\ \mu}_b (x)=\delta^a_b$), $\nabla_\mu(x)=\partial_\mu+\Gamma_\mu(x)$ is the covariant derivative, $\Gamma_\mu(x)=-\frac{i}{4}\sigma^{ab}\omega_{ab\mu}(x)=-\frac{i}{4}\sigma^{ab}e^{\ \nu}_a(x) e_{b\nu}(x)_{;\mu}$ is the spin connection (in fact, $\omega_{ab\mu}(x)$ would be the spin connection, while $\Gamma_\mu(x)$ would be the spinor affine connection or spinorial connection \cite{Carvalho,Cuzinatto,Bragança,O1,O2,O3,O4,O5,O6,JHEP,JCAP}), $\sigma^{ab}=\frac{i}{2}[\gamma^a,\gamma^b]$ are the generators of the Lorentz transformations in a locally Minkowskian frame, $\gamma^5 (x)=-\frac{i}{4!}E^{\mu\nu\alpha\beta}(x)\gamma_\mu (x)\gamma_\nu (x)\gamma_\alpha (x) \gamma_\beta (x)$ is the curved fifth gamma matrix, $E^{\mu\nu\alpha\beta}(x)=\frac{1}{\sqrt{-g}}\varepsilon^{\mu\nu\alpha\beta}$ is the covariant antisymmetric tensor in curved space-time (with $\varepsilon^{txyz}=\varepsilon^{0123}=+1$), $g$ is the determinant of the metric (i.e., $g=$ det$(g_{\mu\nu})$), and $J^\mu$ is the effective external
current of background fermions (background matter or matter potential), with $J^0=V=-\frac{G_F}{\sqrt{2}}n_n\neq 0$ (in fact, it should be $J^t=V\neq 0$), being $G_F=1.17\times 10^{-5}$GeV$^{-2}$ and $n_n$ the Fermi constant and the neutron density, respectively. So, unlike $\gamma_\mu(x)$, for Dvornikov \cite{Dvornikov} (also in \cite{JHEP2,JCAP2}), the effective potential do not depend on the vierbein vectors, i.e., $J^\mu\neq J^\mu(x)=e^{\ \mu}_{a}(x)J^a$.

According to Dvornikov \cite{Dvornikov}, the treatment of the neutrino evolution in a linearly accelerated frame (Rindler space-time) can be given by the following interval (relativistic line element in three-dimensional Cartesian coordinates)
\begin{equation}\label{interval1}
ds^2=g_{\mu\nu}dx^\mu dx^\nu=a^2z^2dt^2-dx^2-dy^2-dz^2,
\end{equation}
where $g_{\mu\nu}=g_{\mu\nu}(x)$ is the metric tensor of the effective gravitational field and $a$ is the proper acceleration of matter. So, one can check that the metric tensor in Eq. \eqref{interval1} can be diagonalized using the following vierbein vectors \cite{Dvornikov}
\begin{eqnarray}\label{vierbein1}
&& e_0^{\ \mu}(x)=\left(\frac{1}{az},0,0,0\right),
\nonumber\\
&& e_1^{\ \mu}(x)=\left(0,1,0,0\right),
\nonumber\\
&& e_2^{\ \mu}(x)=\left(0,0,1,0\right),
\nonumber\\
&& e_3^{\ \mu}(x)=(0,0,0,1).
\end{eqnarray}

With this, the only nonzero component of the connection one-form $\omega_{ab}=\omega_{ab\mu}dx^\mu$ is given as follows \cite{Dvornikov}
\begin{equation}
\omega_{01\mu}=-\omega_{10\mu}=(a,0,0,0),
\end{equation}
where implies that $i\gamma^\mu (x)\Gamma_\mu(x)=\frac{i}{2z}\gamma^3$ \cite{Dvornikov}.

Therefore, with all this, Dvornikov \cite{Dvornikov} rewrote Eq. \eqref{1} in the following equation (with $J^\mu=(V,0,0,0)$ and $\gamma^5(x)=-\frac{i}{az}[\gamma_t (x)\gamma_x (x)\gamma_y(x)\gamma_z(z)]=-\frac{i}{az}[az\gamma_0\gamma_1\gamma_1\gamma_3]=-i[(+\gamma^0)(-\gamma^1)(-\gamma^2)(-\gamma^3)]=i\gamma^0\gamma^1\gamma^2\gamma^3=\gamma^5$, i.e., both $J^\mu$ and $\gamma^5(x)$ do not depends on vierbein vectors or metric, or better, do not depends of the Rindler space-time)
\begin{equation}\label{2}
\left[i\gamma^0\frac{\partial_0}{az}+i\gamma^1 \partial_x+i\gamma^2\partial_y+i\gamma^3\left(\partial_z+\frac{1}{2z}\right)-m\right]\psi=\frac{1}{2}az\gamma^0V(1-\gamma^5)\psi,
\end{equation}
where the flat matrices $\gamma^0$, $\gamma^k$ $(k=1,2,3)$, and $\gamma^5$ are written as (chiral representation)
\begin{equation}\label{matrices}
\gamma^0=\left(
    \begin{array}{cc}
      0\ &  -1 \\
      -1\ & 0 \\
    \end{array}
  \right), \ \  \gamma^k=\left(
    \begin{array}{cc}
      0 & \sigma^k \\
      -\sigma^k & \ 0 \\
    \end{array}
  \right), \ \  \gamma^5=\left(
    \begin{array}{cc}
      1 & \ 0 \\
      0 & -1 \\
    \end{array}
  \right).
\end{equation}

So, considering the following spinor
\begin{equation}\label{spinor}
\psi=\text{exp}(-iEt+ip_x x+ip_y y)\psi_z, 
\end{equation}
with $\psi_z=\psi_z(z)$ being the wave function depending on $z$, Dvornikov \cite{Dvornikov} obtained
\begin{equation}\label{3}
\left[\gamma^a Q_a-m+U\right]\psi_z=0,
\end{equation}
where $Q^a=q^a-q_{\text{eff}}A^{a}_{\text{eff}}$, being $q_{\text{eff}}$ the effective electric charge, $q^a=\left(0,p_x,p_y,-i\partial_z\right)$, and $A^{a}_{\text{eff}}$ is the potential (``vector potential'') of the effective electromagnetic field, given as follows
\begin{equation}\label{4}
A^{a}_{\text{eff}}=\frac{1}{q_{\text{eff}}}\left(\frac{azV}{2}-\frac{E}{az},0,0,\frac{i}{2z}\right),
\end{equation}
where $U=azV\gamma^0\gamma^5/2$ is a type of linear potential at $z$ or simply a linear potential. However, since $a$ has the dimension of the inverse of the length or of $z$ (and $\gamma^0\gamma^5$ is dimensionless), it implies that $U$ also has the dimension of $V$ and, therefore, the dimension of energy (or better, potential energy) \cite{Pal}.

Besides, Dvornikov \cite{Dvornikov} considered the solution of Eq. \eqref{3} in the form: $\psi_z=[\gamma^a Q_a+m-U]\Phi$, where $\Phi=\Phi(z)$ is a new spinor. Unfortunately, Dvornikov \cite{Dvornikov} does not justify why he uses/does this. However, according to several works in the literature, define $\psi_z$ as being the ``Dirac equation with the signs of $m$ and $U$ reversed''(i.e., $m\to -m$ e $U\to -U$), it aims to find a ``quadratic Dirac equation'', that is, find a second-order differential equation without passing directly through the first-order differential equations coupled with the spinor components (such as is done in Refs. \cite{Carvalho,Cuzinatto,Bragança,Guvendi,Vitoria,Bakke,Cunha,O1,O2,O3,O4,O5,O6,JHEP}). In particular, this method for obtaining a ``quadratic Dirac equation'' directly through the linear Dirac equation (i.e., original Dirac equation), has already been used in Refs. \cite{Chen,Gavrilov,Vakarchuk,Peres,Feynman,Auvil,Oliveira1,Oliveira2,Oliveira3,Oliveira4}.

Therefore, using \eqref{4} andhe form of $\psi_z$ (defined just above), Dvornikov \cite{Dvornikov} obtained the following second-order differential equation (or ``quadratic Dirac equation'') for the spinor $\Phi$
\begin{eqnarray}\label{5}
&& \left[\left(\partial_z+\frac{1}{2z}\right)^2+f^2-p^2_\perp+\frac{a^2z^2V^2}{4}-m^2+\frac{f}{z}i\alpha_3-\frac{aV}{2}\left[2zf-i\alpha_3\right]\gamma^5+mazV\gamma^0\gamma^5\right]\Phi=0,
\end{eqnarray}
where $f=f(z)=\left(\frac{E}{az}-\frac{azV}{2}\right)$, $p^2_\perp=p^2_x+p^2_y$, $\alpha_3=\gamma^5\Sigma_3$ and $\Sigma_3=\gamma^0\gamma^3\gamma^5$.

So, according to Dvornikov \cite{Dvornikov}, the solution of Eq. \eqref{5} can be found for ultrarelativistic particles (i.e., in the limit $m\to 0$). With this, it is possible to write $\Phi$ in the form $\Phi=\upsilon\varphi$, where $\varphi=\varphi(z)$ is a scalar function and $\upsilon$ is a constant spinor satisfying $\Sigma_3\upsilon=\sigma\upsilon$ and $\gamma^5\upsilon=\chi\upsilon$, with $\sigma=\pm 1$ and $\chi=\pm 1$ \cite{Dvornikov} (unfortunately, nothing was said about the form of such a spinor, nor about the eigenvalues of $\gamma^0\gamma^5$. Furthermore, to find the eigenvalues, we believe that the eigenvalue equation was used, where $\Lambda\upsilon=\lambda\upsilon$ implies in det$(\Lambda-\lambda)=0$, being $\Lambda=\{\Sigma_3, \gamma^5, \gamma^0\gamma^5\}$). That is, $\sigma=\pm 1$ are the eigenvalues of $\Sigma_3$, while $\chi=\pm 1$ are the eigenvalues of $\gamma^5$ (and the eigenvalues of $\gamma^0\gamma^5$ are $\pm i$, i.e., complex eigenvalues). Besides, Dvornikov \cite{Dvornikov} studied left active neutrinos where $(1+\gamma^5)\psi=0$ and $\chi=+1$. In this way, using a new variable in Eq. \eqref{5} given by $\rho=\vert V\vert az^2$ (i.e., making a change of variable), Dvornikov \cite{Dvornikov} obtained the following equation for $\varphi_\sigma$ (unfortunately, many details were omitted to arrive at this)
\begin{equation}\label{6}
\left[\rho\partial^2_{\rho}+\partial_{\rho}-\frac{\mu^2}{\rho}+\frac{\rho}{4}-\kappa\right]\varphi_\sigma=0,
\end{equation}
where 
\begin{equation}\label{k}
\kappa=\kappa_0+s\kappa_1-i\frac{\sigma s}{4}, \ \ \kappa_0=\frac{p^2_\perp+m^2}{4\vert V\vert a} \ \ \kappa_1=\frac{E}{2a}, \ \ \mu=\frac{1}{4}-i\sigma\kappa_1,
\end{equation}
and $s=$sgn$(V)=-1$ for neutrinos in a neutron matter (since $V=-G_F n_n/\sqrt{2}$).

According to Dvornikov \cite{Dvornikov}, the solutions of Eq. \eqref{6} (Whittaker equation) depend on the Whittaker functions $M_{i\zeta\kappa,\mu}(i\zeta\rho)$ and $W_{i\zeta\kappa,\mu}(i\zeta\rho)$ (with $\zeta=\pm 1$) of the following form: $\varphi_\sigma=\frac{1}{\sqrt{\rho}}M_{i\zeta\kappa,\mu}(i\zeta\rho)$ and $\varphi_\sigma=\frac{1}{\sqrt{\rho}}W_{i\zeta\kappa,\mu}(i\zeta\rho)$. To achieve this, Dvornikov \cite{Dvornikov} used as a basis Ref. \cite{Villalba}, in which the curved Dirac equation was studied in 1+1 de Sitter spacetime and its solutions were also expressed in terms of Whittaker functions. However, in Ref. \cite{Villalba}, the only allowed Whittaker solution is given by $W_{i\zeta\kappa,\mu}(i\zeta\rho)$; consequently, this function can be written as $W_{i\zeta\kappa,\mu}(i\zeta\rho)=\rho^{\mu}e^{-\rho/2}U(1/2-i\zeta\kappa+\mu, 2\mu+1,\rho)$, where $U(1/2-i\zeta\kappa+\mu, 2\mu+1,\rho)$ is the confluent hypergeometric function of the second kind. On the other hand, as we will see in the next section, we will obtain a Wittakker function given by $M$, which will be written in terms of the confluent hypergeometric function of the first kind (or even through the generalized/associated Laguerre polynomials). Before concluding this section, let us make a remark about Dvornikov's differential equation. So, as stated by Dvornikov himself \cite{Dvornikov}, the solution of Eq. \eqref{5} can be found for ultrarelativistic particles ($m\to 0$); consequently, Eq. \eqref{6} should not contain $m$ (or better, $m^2$). However, that is not what happened; that is, is somewhat contradictory. In particular, this contradiction was also made in another paper of yours, given by Ref. \cite{JCAP2}. However, in this reference, one really should have considered $m\to 0$, which would eliminate $\gamma^0\gamma^5$ from the differential equation and, consequently, would avoid obtaining complex energies (that is, bound-state solutions require a quantized and real energy spectrum). So, unlike \cite{JCAP2}, here (and in \cite{Dvornikov}), it is not necessary to do $m\to 0$ since the differential equation is already complex by ``nature'', that is, a consequence of the adopted space-time (in fact, even for $V=0$, Eq. \eqref{5} still remains complex due to the term $iE\alpha_3/az^2$).


\section{The true/correct second-order differential equation generated from the curved Dirac equation for a neutrino interacting with linearly accelerated matter}

According to Refs. \cite{Bandyopadhyay,JCAP}, the true/correct curved Dirac equation for a neutrino interacting with linearly accelerated matter is written as follows
\begin{equation}\label{7}
[i\gamma^\mu (x)\nabla_\mu (x)-m]\psi=\frac{1}{2}J^\mu(x)\gamma_\mu (x)[1-\gamma^5 (x)]\psi,
\end{equation}
where the curved effective external current is given by $J^\mu (x)=e^{\ \mu}_b (x)J^b$ (i.e., depends on the vierbein vectors), being $J^b$ the flat/usual effective external current. However, as a consequence of the orthogonality condition of the vierbein vectors and their inverses, given by $e^a_{\ \mu}(x) e^{\ \mu}_b (x)=\delta^a_b$, we have $J^\mu(x)\gamma_\mu (x)=e^{\ \mu}_b (x)J^be^a_{\ \mu}(x)\gamma_a=e^{\ \mu}_b e^a_{\ \mu}(x)(x)J^b\gamma_a=\delta^a_b J^b\gamma_a=J^a\gamma_a$, i.e., the product $J^\mu(x)\gamma_\mu (x)$ do not depend on vierbein vectors or metric (as well as $\gamma^5 (x)$). Therefore, Eq. \eqref{7} becomes
\begin{equation}\label{7}
[i\gamma^\mu (x)\nabla_\mu (x)-m]\psi=\frac{1}{2}J^a\gamma_a[1-\gamma^5]\psi,
\end{equation}
or better
\begin{equation}\label{8}
[i\gamma^\mu (x)\nabla_\mu (x)-m]\psi=\frac{1}{2}\gamma^0 V[1-\gamma^5]\psi.
\end{equation}

That is, unlike the incorrect Dirac equation (given by Eq. \eqref{2}), the electroweak interaction term (or simply interaction term or even matter term) in the correct Dirac equation (i.e., last term of \eqref{7} or \eqref{8}) does not depend on the spacetime coordinates \cite{Bandyopadhyay,JCAP}. In this way, under the influence (or the effects) of a linearly accelerated matter, Eq. \eqref{8} becomes
\begin{equation}\label{9}
\left[i\gamma^0\frac{\partial_0}{az}+i\gamma^1 \partial_x+i\gamma^2\partial_y+i\gamma^3\left(\partial_z+\frac{1}{2z}\right)-m\right]\psi=\frac{1}{2}\gamma^0 V[1-\gamma^5]\psi,
\end{equation}
or better
\begin{equation}\label{10}
\left[-\gamma^0\left(\frac{V}{2}-\frac{E}{az}\right)-\gamma^1 p_x-\gamma^2p_y+i\gamma^3\left(\partial_z+\frac{1}{2z}\right)-m+\bar{U}\right]\psi_z=0,
\end{equation}
where we define $\bar{U}=\frac{V}{2}\gamma^0\gamma^5$ (a constant potential since $V$ has a dimension of potential energy \cite{Pal}), and we use the spinor \eqref{spinor}. That is, unlike $U$, here, our potential $\bar{U}$ does not depend on $z$ (and not even of $a$). In other words (or for simplicity), our potential is half of the effective external current (or matter potential).

In tensor or index notation, Eq. \eqref{10} can be written as
\begin{equation}\label{11}
\left[\gamma^a\bar{Q}_a-m+\bar{U}\right]\psi_z=0,
\end{equation}
where $\bar{Q}^a=q^a-q_{\text{eff}}\bar{A}^{a}_{\text{eff}}$ (and $\bar{Q}_a=q_a-q_{\text{eff}}\bar{A}_{a}^{\text{eff}}$), being $q^a=\left(0,p_x,p_y,-i\partial_z\right)$ (and $q_a=\left(0,-p_x,-p_y,+i\partial_z\right)$ due to the metric signature, which is $(+,-,-,-)$), and $\bar{A}^{a}_{\text{eff}}$ is defined as follows
\begin{equation}\label{12}
\bar{A}^{a}_{\text{eff}}=\frac{1}{q_{\text{eff}}}\left(\frac{V}{2}-\frac{E}{az},0,0,\frac{i}{2z}\right) \ \ \leftrightarrow \ \ \bar{A}_{a}^{\text{eff}}=\frac{1}{q_{\text{eff}}}\left(\frac{V}{2}-\frac{E}{az},0,0,-\frac{i}{2z}\right).
\end{equation}

As we see above, unlike the incorrect equation, here, the only part of the zero/time component of our ``vector potential'' $\bar{A}_{\text{eff}}^a$ that depends on $z$ is the term $-E/az$. In fact, as we will see below, this will result in a second-order differential equation very different from the one obtained by Dvornikov \cite{Dvornikov}. So, defining $\psi_z=\left[\gamma^b\bar{Q}_b+m-\bar{U}\right]\bar{\Phi}$, we have
\begin{equation}\label{13}
\left[\gamma^a\bar{Q}_a-m+\bar{U}\right]\left[\gamma^b\bar{Q}_b+m-\bar{U}\right]\bar{\Phi}=0,
\end{equation}
where implies
\begin{align}\label{14}
\left[\gamma^a\bar{Q}_a-m+\bar{U}\right]\left[\gamma^b\bar{Q}_b+m-\bar{U}\right]&=\gamma^a\bar{Q}_a\gamma^b\bar{Q}_b+\gamma^a\bar{Q}_a m-\gamma^a\bar{Q}_a\bar{U}-m\gamma^b\bar{Q}_b-m^2+m\bar{U}+\bar{U}\gamma^b\bar{Q}_b+\bar{U}m-\bar{U}\bar{U},
\nonumber\\
&=\gamma^a\gamma^b\bar{Q}_a\bar{Q}_b+\bar{Q}_a[\bar{U}\gamma^a-\gamma^a\bar{U}]-m^2+2m\bar{U}-\bar{U}\bar{U},
\nonumber\\
&=\gamma^a\gamma^b\bar{Q}_a\bar{Q}_b+\bar{Q}_0[\bar{U}\gamma^0-\gamma^0\bar{U}]-m^2+mV\gamma^0\gamma^5-\frac{V}{2}\gamma^0\gamma^5\frac{V}{2}\gamma^0\gamma^5,
\nonumber\\
&=\gamma^a\gamma^b\bar{Q}_a\bar{Q}_b+\bar{Q}_0\left[\frac{V}{2}\gamma^0\gamma^5\gamma^0-\gamma^0\frac{V}{2}\gamma^0\gamma^5\right]-m^2+mV\gamma^0\gamma^5-\frac{V^2}{4}\gamma^0\gamma^5\gamma^0\gamma^5,
\nonumber\\
&=\gamma^a\gamma^b\bar{Q}_a\bar{Q}_b+\bar{Q}_0\left[-\frac{V}{2}\gamma^0\gamma^0\gamma^5-\gamma^0\gamma^0\frac{V}{2}\gamma^5\right]-m^2+mV\gamma^0\gamma^5+\frac{V^2}{4}\gamma^0\gamma^0\gamma^5\gamma^5,
\nonumber\\
&=\gamma^a\gamma^b\bar{Q}_a\bar{Q}_b+\bar{Q}_0\left[-\frac{V}{2}\gamma^5-\frac{V}{2}\gamma^5\right]-m^2+mV\gamma^0\gamma^5+\frac{V^2}{4},
\nonumber\\
&=\gamma^a\gamma^b\bar{Q}_a\bar{Q}_b-V\bar{Q}_0\gamma^5-m^2+mV\gamma^0\gamma^5+\frac{V^2}{4},
\nonumber\\
&=\gamma^a\gamma^b\bar{Q}_a\bar{Q}_b+V\left(\frac{V}{2}-\frac{E}{az}\right)\gamma^5-m^2+mV\gamma^0\gamma^5+\frac{V^2}{4},
\end{align}
where we use the fact that $\gamma^0\gamma^5=-\gamma^5\gamma^0$, $\gamma^0\gamma^0=(\gamma^0)^2=\gamma^5\gamma^5=(\gamma^5)^2=1$, and $\bar{Q}_0=q_0-q_{\text{eff}}\bar{A}^{\text{eff}}_0=-\left(\frac{V}{2}-\frac{E}{az}\right)$.

Now, we need to develop (``open the indices'') the term $\gamma^a\gamma^b\bar{Q}_a\bar{Q}_b$. So, we have
\begin{align}\label{15}
\gamma^a\gamma^b\bar{Q}_a\bar{Q}_b&=\gamma^0\gamma^b\bar{Q}_0\bar{Q}_b+\gamma^1\gamma^b\bar{Q}_1\bar{Q}_b+\gamma^2\gamma^b\bar{Q}_2\bar{Q}_b+\gamma^3\gamma^b\bar{Q}_3\bar{Q}_b,
\nonumber\\
&=\gamma^0\gamma^0\bar{Q}_0\bar{Q}_0+\gamma^0\gamma^1\bar{Q}_0\bar{Q}_1+\gamma^0\gamma^2\bar{Q}_0\bar{Q}_2+\gamma^0\gamma^3\bar{Q}_0\bar{Q}_3,
\nonumber\\
&+\gamma^1\gamma^0\bar{Q}_1\bar{Q}_0+\gamma^1\gamma^1\bar{Q}_1\bar{Q}_1+\gamma^1\gamma^2\bar{Q}_1\bar{Q}_2+\gamma^1\gamma^3\bar{Q}_1\bar{Q}_3,
\nonumber\\
&+\gamma^2\gamma^0\bar{Q}_2\bar{Q}_0+\gamma^2\gamma^1\bar{Q}_2\bar{Q}_1+\gamma^2\gamma^2\bar{Q}_2\bar{Q}_2+\gamma^2\gamma^3\bar{Q}_2\bar{Q}_3,
\nonumber\\
&+\gamma^3\gamma^0\bar{Q}_3\bar{Q}_0+\gamma^3\gamma^1\bar{Q}_3\bar{Q}_1+\gamma^3\gamma^2\bar{Q}_3\bar{Q}_2+\gamma^3\gamma^3\bar{Q}_3\bar{Q}_3,
\nonumber\\
&=\bar{Q}_0^2-\bar{Q}_1^2-\bar{Q}_2^2-\bar{Q}_3^2+\gamma^0\gamma^1[\bar{Q}_0\bar{Q}_1-\bar{Q}_1\bar{Q}_0]+\gamma^0\gamma^2[\bar{Q}_0\bar{Q}_2-\bar{Q}_2\bar{Q}_0]+\gamma^0\gamma^3[\bar{Q}_0\bar{Q}_3-\bar{Q}_3\bar{Q}_0],
\nonumber\\
&+\gamma^1\gamma^2[\bar{Q}_1\bar{Q}_2-\bar{Q}_2\bar{Q}_1]+\gamma^1\gamma^3[\bar{Q}_1\bar{Q}_3-\bar{Q}_3\bar{Q}_1]+\gamma^2\gamma^3[\bar{Q}_2\bar{Q}_3-\bar{Q}_3\bar{Q}_2],
\nonumber\\
&=\bar{Q}_0^2-\bar{Q}_1^2-\bar{Q}_2^2-\bar{Q}_3^2+\gamma^0\gamma^3[\bar{Q}_0\bar{Q}_3-\bar{Q}_3\bar{Q}_0],
\end{align}
where we use the fact that $\gamma^i\gamma^i=(\gamma^i)^2=-1$ ($i=1,2,3$), $\gamma^0\gamma^i=-\gamma^i\gamma^0$, $\bar{Q}_0\bar{Q}_{\bar{i}}=\bar{Q}_{\bar{i}}\bar{Q}_0$ ($\bar{i}=1,2$), $\bar{Q}_{\bar{j}}\bar{Q}_{\bar{k}}=\bar{Q}_{\bar{k}}\bar{Q}_{\bar{j}}$ ($\bar{j},\bar{k}=1,2,3$), and $\bar{Q}_0\bar{Q}_3\neq \bar{Q}_3\bar{Q}_0$. Continuing, we have
\begin{align}\label{16}
[\bar{Q}_0\bar{Q}_3-\bar{Q}_3\bar{Q}_0]\bar{\Phi}&=\left(-\left(\frac{V}{2}-\frac{E}{az}\right)\right)\left(i\partial_z+\frac{i}{2z}\right)\bar{\Phi}-\left(i\partial_z+\frac{i}{2z}\right)\left(-\left(\frac{V}{2}-\frac{E}{az}\right)\right)\bar{\Phi},
\nonumber\\
&=-\left(\frac{V}{2}-\frac{E}{az}\right)\left(i\partial_z\bar{\Phi}+\frac{i}{2z}\bar{\Phi}\right)+\left(i\partial_z+\frac{i}{2z}\right)\left(\frac{V}{2}\bar{\Phi}-E\frac{\bar{\Phi}}{az}\right),
\nonumber\\
&=-\frac{V}{2}\left(i\partial_z\bar{\Phi}+\frac{i}{2z}\bar{\Phi}\right)+\frac{E}{az}\left(i\partial_z\bar{\Phi}+\frac{i}{2z}\bar{\Phi}\right)+i\partial_z\left(\frac{V}{2}\bar{\Phi}-E\frac{\bar{\Phi}}{az}\right)+\frac{i}{2z}\left(\frac{V}{2}\bar{\Phi}-E\frac{\bar{\Phi}}{az}\right),
\nonumber\\
&=-\frac{V}{2}i\partial_z\bar{\Phi}-\frac{V}{2}\frac{i}{2z}\bar{\Phi}+\frac{iE}{az}\partial_z\bar{\Phi}+\frac{E}{az}\frac{i}{2z}\bar{\Phi}+\frac{V}{2}i\partial_z\bar{\Phi}-\frac{iE}{a}\partial_z\left(\frac{\bar{\Phi}}{z}\right)+\frac{i}{2z}\frac{V}{2}\bar{\Phi}-\frac{iE}{2z}\frac{\bar{\Phi}}{az},
\nonumber\\
&=-\frac{V}{2}i\partial_z\bar{\Phi}-\frac{V}{2}\frac{i}{2z}\bar{\Phi}+\frac{iE}{az}\partial_z\bar{\Phi}+\frac{E}{az}\frac{i}{2z}\bar{\Phi}+\frac{V}{2}i\partial_z\bar{\Phi}-\frac{iE}{a}\left(\frac{1}{z}\partial_z\bar{\Phi}-\frac{\bar{\Phi}}{z^2}\right)+\frac{i}{2z}\frac{V}{2}\bar{\Phi}-\frac{iE}{2z}\frac{\bar{\Phi}}{az},
\nonumber\\
&=-\frac{V}{2}i\partial_z\bar{\Phi}-\frac{V}{2}\frac{i}{2z}\bar{\Phi}+\frac{iE}{az}\partial_z\bar{\Phi}+\frac{E}{az}\frac{i}{2z}\bar{\Phi}+\frac{V}{2}i\partial_z\bar{\Phi}-\frac{iE}{a}\frac{1}{z}\partial_z\bar{\Phi}+\frac{iE}{a}\frac{\bar{\Phi}}{z^2}+\frac{i}{2z}\frac{V}{2}\bar{\Phi}-\frac{iE}{2z}\frac{\bar{\Phi}}{az},
\nonumber\\
&=\frac{E}{az}\frac{i}{2z}\bar{\Phi}+\frac{iE}{a}\frac{\bar{\Phi}}{z^2}-\frac{iE}{2z}\frac{\bar{\Phi}}{az},
\nonumber\\
&=\frac{iE}{az^2}\bar{\Phi}.
\end{align}

Therefore, knowing that $\bar{Q}_0=-(V/2-E/az)$, $\bar{Q}_1=-p_x$, $\bar{Q}_2=-p_y$, and $\bar{Q}_3=(i\partial_z+i/2z)$, we have
\begin{align}\label{17}
\gamma^a\gamma^b\bar{Q}_a\bar{Q}_b&=\bar{Q}_0^2-\bar{Q}_1^2-\bar{Q}_2^2-\bar{Q}_3^2+\gamma^0\gamma^3[\bar{Q}_0\bar{Q}_3-\bar{Q}_3\bar{Q}_0],
\nonumber\\
&=\left(\frac{V}{2}-\frac{E}{az}\right)^2-[p_x^2+p_y^2]-\left(i\partial_z+\frac{i}{2z}\right)^2+\frac{iE}{az^2}\gamma^0\gamma^3,
\end{align}
or better
\begin{align}\label{18}
\gamma^a\gamma^b\bar{Q}_a\bar{Q}_b&=\left(\frac{d}{dz}+\frac{1}{2z}\right)^2+\left(\frac{E}{az}-\frac{V}{2}\right)^2-p^2_\perp+\frac{iE}{az^2}\alpha^3,
\nonumber\\
&=\frac{d^2}{dz^2}+\frac{1}{z}\frac{d}{dz}-\frac{1}{4z^2}+\frac{E^2}{a^2z^2}-\frac{EV}{az}+\frac{V^2}{4}-p^2_\perp+\frac{iE}{az^2}\alpha^3,
\nonumber\\
&=\frac{d^2}{dz^2}+\frac{1}{z}\frac{d}{dz}-\frac{\frac{1}{4}-\frac{E^2}{a^2}-\frac{iE}{a}\alpha^3}{z^2}-\frac{EV}{az}+\frac{V^2}{4}-p^2_\perp,
\nonumber\\
&=\frac{d^2}{dz^2}+\frac{1}{z}\frac{d}{dz}-\frac{(\frac{1}{2}-\frac{iE}{a}\alpha^3)^2}{z^2}-\frac{EV}{az}+\frac{V^2}{4}-p^2_\perp,
\end{align}
where we use $p^2_\perp=p_x^2+p_y^2$, and $\alpha^3=\gamma^0\gamma^3=$diag$(\sigma^3,-\sigma^3)$ (with $(\alpha^3)^2=1$).

Consequently, Eq. \eqref{13} becomes (i.e., the true/correct second-order differential equation for the neutrino)
\begin{equation}\label{19}
\left[\frac{d^2}{dz^2}+\frac{1}{z}\frac{d}{dz}-\frac{(\frac{1}{2}-\frac{iE}{a}\alpha^3)^2}{z^2}-\frac{EV}{az}+\frac{V^2}{4}-p^2_\perp+V\left(\frac{V}{2}-\frac{E}{az}\right)\gamma^5-m^2+mV\gamma^0\gamma^5+\frac{V^2}{4}\right]\bar{\Phi}=0,
\end{equation}
or better
\begin{equation}\label{20}
\left[\frac{d^2}{dz^2}+\frac{1}{z}\frac{d}{dz}-\frac{(\frac{1}{2}-\frac{iE}{a}\alpha^3)^2}{z^2}-\frac{EV}{az}(1+\gamma^5)-p^2_\perp-m^2+mV\gamma^0\gamma^5+\frac{V^2}{2}(1+\gamma^5)\right]\bar{\Phi}=0.
\end{equation}

It is important to highlight that, here, $(1+\gamma^5)$ has nothing to do with the one in the previous section; that is, it appeared here only as a matter of organizing some terms of the equation. So, written $\bar{\Phi}$ as $\bar{\Phi}=\upsilon\bar{\varphi}$, where $\alpha^3\upsilon=\sigma\upsilon$ ($\sigma=\pm 1$), $\gamma^5\upsilon=\chi\upsilon=+\upsilon$ (in fact, if $\gamma^5\upsilon=-\upsilon$ were used, the equation above will not depend in any way on $V$), and $\gamma^0\gamma^5\upsilon=is\upsilon$ ($s=\pm 1$), we obtain the following equation for $\bar{\varphi}$ (or better, $\bar{\varphi}_{\sigma,s}$)
\begin{equation}\label{21}
\left[\frac{d^2}{dz^2}+\frac{1}{z}\frac{d}{dz}-\frac{\bar{\mu}^2}{z^2}+\frac{2E\vert V\vert}{az}-\kappa^2\right]\bar{\varphi}_{\sigma,s}=0,
\end{equation}
where $\kappa$ and $\bar{\mu}$ are two complex parameters, defined as follows
\begin{equation}
\kappa=\kappa_s\equiv\sqrt{p^2_\perp+m^2+ism\vert V\vert-\vert V\vert^2}, \ \ \bar{\mu}=\bar{\mu}_\sigma\equiv\frac{1}{2}-\frac{i\sigma E}{a},
\end{equation}
where we use the fact that $V=-\vert V \vert$. 

With the purpose of solving Eq. \eqref{21}, let us first consider for simplicity that $0\leq z<\infty$ (or $z^2\to \vert z\vert^2$ in \eqref{interval1}), i.e., $z$ behaves like a ``spherical radial coordinate'' (this will be justified soon). So, writing $\bar{\varphi}_{\sigma,s}(z)$ as \cite{Bragança}
\begin{equation}\label{F}
\bar{\varphi}_{\sigma,s}(z)=\frac{F_{\sigma,s}(z)}{\sqrt{z}},
\end{equation}
we obtain, as a consequence (i.e., a change of function), the following equation for the function $F_{\sigma,s}(z)$
\begin{equation}\label{22}
\left[\frac{d^2}{dz^2}+\frac{\frac{1}{4}-\bar{\mu}^2}{z^2}+\frac{2E\vert V\vert}{az}-\kappa^2\right]F_{\sigma,s}(z)=0.
\end{equation}

Now, starting from the fact that Eq. \eqref{22} has the form of the hydrogen atom equation (i.e., \eqref{22} is a hydrogen atom-type equation whose ``potential'' is given by $V_{Rindler}(z)=V_{neutrino}(z)=-\frac{2E\vert V\vert}{az}$, with $a>0$ and $E>0$, such as occurs in the case of the relativistic hydrogen atom), we can then use a variable given by $\rho=2\kappa z$ (or better, $z=\rho/2\kappa$). Therefore, by making a change of variable, Eq. \eqref{22} becomes
\begin{equation}\label{23}
\left[\frac{d^2}{d\rho^2}+\frac{\frac{1}{4}-\bar{\mu}^2}{\rho^2}+\frac{\bar{\kappa}}{\rho}-\frac{1}{4}\right]F_{\sigma,s}(\rho)=0,
\end{equation}
where we define $\bar{\kappa}\equiv\frac{E\vert V\vert}{a\kappa}$ (that is, it is also a complex parameter due to $\kappa$).

According to Refs. \cite{Bragança,Arfken}, Eq. \eqref{23} is the well-known Whittaker equation (modeled by complex parameters $\bar{\kappa}$ and $\bar{\mu}$) and $F_{\sigma,s}(\rho)$ is the Whittaker function, which can be written in terms of the confluent hypergeometric function of the first kind $_{1}F_{1}(\rho)$ in the following form
\begin{equation}\label{Whittaker1}
F_{\sigma,s}(\rho)=M_{\bar{\kappa},\bar{\mu}}(\rho)=\rho^{1/2+\bar{\mu}}{e^{-\rho/2}}_{1}F_{1}\left(\bar{\mu}-\bar{\kappa}+\frac{1}{2},2\bar{\mu}+1;\rho\right),
\end{equation}
or better
\begin{equation}\label{Whittaker2}
F_{\sigma,s}(\rho)=M_{\bar{\kappa},\bar{\mu}}(\rho)=C\rho^{1/2+\bar{\mu}}{e^{-\rho/2}}_{1}F_{1}\left(\bar{\mu}-\bar{\kappa}+\frac{1}{2},2\bar{\mu}+1;\rho\right),
\end{equation}
where $C=C_{\sigma,s}$ is a normalization constant (and it is complex, of course). On the other hand, we can also write $F_{\sigma,s}(\rho)$ in terms of the associated Laguerre polynomials $L^\gamma_n (x)$ using the relation $_{1}F_{1}(-n,\gamma+1,x)=\frac{n!\gamma!}{(n+\gamma!)}L^\gamma_n (x)$ \cite{Bragança,Arfken}. In this case, we have: $F_{\sigma,s}(\rho)=\bar{C}\rho^{1/2+\bar{\mu}}e^{-\rho/2}L^{2\bar{\mu}}_{\bar{\kappa}-\bar{\mu}-1/2}(\rho)$, where $\bar{C}$ would be a new constant, given by $\bar{C}=\frac{(\bar{\kappa}-\bar{\mu}-1/2)!(2\bar{\mu})!}{(\bar{\kappa}+\bar{\mu}-1/2)}C$. In particular, among some works whose solutions of the Dirac equation are written directly (or transformed) in terms of the associated Laguerre polynomials, we can mention Refs. \cite{Chen,O3,O4,O5,O6,JHEP,JHEP2,Auvil,Oliveira2,Oliveira3,Oliveira4,Villalba2}.

Therefore, the true/correct Whittaker equation for a Dirac neutrino interacting with uniformly accelerated matter is given by Eq. \eqref{23}, where the true/correct Whittaker function is given by the function \eqref{Whittaker2}. So, using $z=\rho/2\kappa$, the function \eqref{F} becomes (i.e., the true/correct scalar function of the Dirac spinor written in terms of $\rho$)
\begin{equation}\label{F2}
\bar{\varphi}_{\sigma,s}(\rho)=\frac{M_{\bar{\kappa},\bar{\mu}}(\rho)}{\sqrt{\rho/2\kappa}}=\frac{\bar{M}_{\bar{\kappa},\bar{\mu}}(\rho)}{\sqrt{\rho}}, \ \ (\bar{M}_{\bar{\kappa},\bar{\mu}}(\rho)\equiv\sqrt{2\kappa}M_{\bar{\kappa},\bar{\mu}}(\rho)).
\end{equation}


\section{Final remarks}

In the present comment, we show that the fundamental equation worked by Dvornikov \cite{Dvornikov} in his paper, which is the Dirac equation for a massive neutrino interacting with linearly accelerated matter (i.e., in the Rindler space-time), is incorrect. In particular, Dvornikov \cite{Dvornikov} incorrectly wrote/defined the effective external current of background fermions (background matter or matter potential) in a curved space-time. That is, such an effective current should also depend on the vierbein vectors; however, this was not what happened. In other words, Dvornikov \cite{Dvornikov} wrote/defined such an effective current in a flat space-time, which is a mistake. Consequently, the second-order differential equation (generated through the quadratic Dirac equation) in your paper is incorrect, where such an equation is given by the Whittaker equation. So, since the solutions of such a differential equation (whose solutions are the Whittaker functions) are the basis for its results, it implies that such results are also incorrect. In this way, starting from the true/correct Dirac equation with an effective external current in a curved space-time, we obtain in detail the second-order differential equation (also a Whittaker equation) and its solutions for a neutrino interacting with linearly accelerated matter.

\section*{Acknowledgments}

\hspace{0.5cm}

The author would like to thank the Conselho Nacional de Desenvolvimento Cient\'{\i}fico e Tecnol\'{o}gico (CNPq) for financial support.

\section*{Data availability statement}

\hspace{0.5cm} This manuscript has no associated data or the data will not be deposited. [Author’ comment: There is no data associated with this manuscript or no data has been used to prepare it.]


\begin{thebibliography}{99}
\section*{References}

\bibitem{Dvornikov} M. Dvornikov, JHEP {\bf 2015}, 151 (2015). {\color{blue}\url{https://doi.org/10.1007/JHEP08(2015)151}. {\href{https://arxiv.org/abs/1507.01174}{arXiv:1507.01174}}}

\bibitem{Carvalho} J. Carvalho, C. Furtado, and F. Moraes, Phys. Rev. A {\bf 84}, 032109 (2011). {\color{blue}\url{https://doi.org/10.1103/PhysRevA.84.032109}}

\bibitem{Cuzinatto} R. R. Cuzinatto, M. de Montigny, and P. J. Pompeia, Gen. Relativ. Gravit. {\bf 51}, 107 (2019). {\color{blue}\url{https://doi.org/10.1007/s10714-019-2593-3}. {\href{https://arxiv.org/abs/1909.00904}{arXiv:1909.00904}}}

\bibitem{Bragança} E. A. F. Bragança, R. L. L. Vitória, H. Belich, and E. B. de Mello, Eur. Phys. J. C {\bf 80}, 206 (2020). {\color{blue}\url{https://doi.org/10.1140/epjc/s10052-020-7774-4}. {\href{https://arxiv.org/abs/1909.00037v2}{arXiv:1909.00037v2}}}

\bibitem{Chen} H. L. Chen, K. Fukushima, X. G. Huang, and K. Mameda, Phys. Rev. D {\bf 93}, 104052 (2016). {\color{blue}\url{https://doi.org/10.1103/PhysRevD.93.104052}. {\href{https://arxiv.org/abs/1512.08974v2}{arXiv:1512.08974v2}}}

\bibitem{Guvendi} A. Guvendi, and Y. Sucu, Phys. Lett. B {\bf 811}, 135960 (2020). {\color{blue}\url{https://doi.org/10.1016/j.physletb.2020.135960}}

\bibitem{Vitoria} R. L. L. Vitória, and K. A. T. da Silva, Int. J. Theor. Phys {\bf 63}, 246 (2024). {\color{blue}\url{https://doi.org/10.1007/s10773-024-05788-4}}

\bibitem{Bakke} K. Bakke, and C. Furtado, Phys. Rev. D {\bf 82}, 084025 (2010). {\color{blue}\url{https://doi.org/10.1103/PhysRevD.82.084025}}

\bibitem{Cunha} M. M. Cunha, H. S. Dias, and E. O. Silva, Phys. Rev. D {\bf 102}, 105020 (2020). {\color{blue}\url{https://doi.org/10.1103/PhysRevD.102.105020}. {\href{https://arxiv.org/abs/2007.08699}{arXiv:2007.08699}}}

\bibitem{O1} R. R. S. Oliveira, Gen. Relativ. Gravit. {\bf 51}, 120 (2019). {\color{blue}\url{https://doi.org/10.1007/s10714-019-2606-2}. {\href{https://arxiv.org/abs/1907.00054}{arXiv:1907.00054}}}

\bibitem{O2} R. R. S. Oliveira, Eur. Phys. J. C {\bf 79}, 725 (2019). {\color{blue}\url{https://doi.org/10.1140/epjc/s10052-019-7237-y}}

\bibitem{O3} R. R. S. Oliveira, Gen. Relativ. Gravit. 52, {\bf 88} (2020). {\color{blue}\url{https://doi.org/10.1007/s10714-020-02743-6}. {\href{https://arxiv.org/abs/1906.07369}{arXiv:1906.07369}}}

\bibitem{O4} R. R. S. Oliveira, G. Alencar, and R. R. Landim, Gen. Relativ. Gravit. {\bf 55}, 15 (2023). {\color{blue}\url{https://doi.org/10.1007/s10714-022-03057-5}. {\href{https://arxiv.org/abs/2204.06057}{arXiv:2204.06057}}}

\bibitem{O5} R. R. S. Oliveira, Gen. Relativ. Gravit. {\bf 56}, 30 (2024). {\color{blue}\url{https://doi.org/10.1007/s10714-024-03209-9}. {\href{https://arxiv.org/abs/2402.15720}{arXiv:2402.15720}}}

\bibitem{O6} R. R. S. Oliveira, Class. Quantum Grav. {\bf 41}, 175017 (2024).
{\color{blue}\url{https://doi.org/10.1088/1361-6382/ad69f5}. {\href{https://arxiv.org/abs/2405.11334}{arXiv:2405.11334}}}

\bibitem{JHEP} R. R. S. Oliveira, JHEP {\bf 2025}, 85 (2025). {\color{blue}\url{https://doi.org/10.1007/JHEP01(2025)085}. {\href{https://arxiv.org/abs/2411.04338}{arXiv:2411.04338}}}

\bibitem{JCAP} R. R. S. Oliveira, JCAP {\bf 9}, 70 (2025). {\color{blue}\url{https://doi.org/10.1088/1475-7516/2025/09/070}. {\href{https://arxiv.org/abs/2502.05977v2}{arXiv:2502.05977v2}}}

\bibitem{Bandyopadhyay} S. Bandyopadhyay, and G. M. Hossain, Phys. Rev. D {\bf 111}, 065009 (2025). {\color{blue}\url{https://doi.org/10.1103/PhysRevD.111.065009}. {\href{https://arxiv.org/abs/2409.16232}{arXiv:2409.16232}}}

\bibitem{JHEP2} M. Dvornikov, JHEP {\bf 2014}, 1-15 (2014). {\color{blue}\url{https://doi.org/10.1007/JHEP10(2014)053}. {\href{https://arxiv.org/abs/1408.2735}{arXiv:1408.2735}}}

\bibitem{JCAP2} M. Dvornikov, JCAP {\bf 2015}, 037 (2015). {\color{blue}\url{https://doi.org/10.1088/1475-7516/2015/05/037}. {\href{https://arxiv.org/abs/1503.00608}{arXiv:1503.00608}}}

\bibitem{Pal} P. B. Pal, Int. J. Mod. Phys. A {\bf 7}, 5387–5459 (1992). {\color{blue}\url{https://doi.org/10.1142/S0217751X92002465}}

\bibitem{Gavrilov} S. P. Gavrilov, and D. M. Gitman, Phys. Rev. D {\bf 53}, 7162 (1996). {\color{blue}\url{https://doi.org/10.1103/PhysRevD.53.7162}. {\href{https://arxiv.org/abs/hep-th/9603152}{arXiv:hep-th/9603152}}}

\bibitem{Vakarchuk} I. O. Vakarchuk, J. Phys. A {\bf 38}, 4727 (2005). {\color{blue}\url{https://doi.org/10.1088/0305-4470/38/21/016}. {\href{https://arxiv.org/abs/quant-ph/0502105}{arXiv:quant-ph/0502105}}}

\bibitem{Peres} N. M. R. Peres, A. H. Castro Neto, and F. Guinea, Phys. Rev. B {\bf 73}, 241403 (2006). {\color{blue}\url{https://doi.org/10.1103/PhysRevB.73.241403}. {\href{https://arxiv.org/abs/cond-mat/0603771}{arXiv:cond-mat/0603771}}}

\bibitem{Feynman} R. P. Feynman, and M. Gell-Mann, Phys. Rev. {\bf 109}, 193 (1958). {\color{blue}\url{https://doi.org/10.1103/PhysRev.109.193}}

\bibitem{Auvil} P. R. Auvil, and L. M. Brown, Am. J. Phys. {\bf 46}, 679 (1978). {\color{blue}\url{https://doi.org/10.1119/1.11231}}

\bibitem{Oliveira1} R. R. S. Oliveira, R. V. Maluf, and C. A. S. Almeida, Ann. Phys. {\bf 400}, 1-8 (2019). {\color{blue}\url{https://doi.org/10.1016/j.aop.2018.11.005}. {\href{https://arxiv.org/abs/1809.03801}{arXiv:1809.03801}}}

\bibitem{Oliveira2} R. R. S. Oliveira, A. A. Araújo Filho, R. V. Maluf, and C. A. S. Almeida, J. Phys. A: Math. Theor. {\bf 53}, 045304 (2020). {\color{blue}\url{https://doi.org/10.1088/1751-8121/ab5cfb}. {\href{https://arxiv.org/abs/1812.07756}{arXiv:1812.07756}}}

\bibitem{Oliveira3} R. R. de Sousa Oliveira, G. Alencar and R.R. Landim, Phys. Scripta {\bf 99}, 035226 (2024). {\color{blue}\url{https://doi.org/10.1088/1402-4896/ad25b3}. {\href{https://arxiv.org/abs/2211.09592}{arXiv:2211.09592}}}

\bibitem{Oliveira4} R. R. S. Oliveira, Int. J. Theor. Phys. {\bf 64}, 38 (2025). {\color{blue}\url{https://doi.org/10.1007/s10773-025-05901-1}. {\href{https://arxiv.org/abs/2405.16300v2}{arXiv:2405.16300v2}}}

\bibitem{Villalba} V. M. Villalba, Phys. Rev. D {\bf 52}, 3742 (1995). 
{\color{blue}\url{https://doi.org/10.1103/PhysRevD.52.3742}}

\bibitem{Arfken} G. B. Arfken, and  H. J. Weber, {\it Mathematical Methods for Physicists}, sixth edition (Elsevier Academic Press, New York, 2005).

\bibitem{Villalba2} V. M. Villalba, Phys. Rev. A {\bf 49}, 586 (1994). {\color{blue}\url{https://doi.org/10.1103/PhysRevA.49.586}}

\end{thebibliography}
\end{document}